%% file: iclr2026_conference.tex
\documentclass{article} 
\usepackage{iclr2027_conference,times}

\input{math_commands.tex}

\usepackage{multirow}
\usepackage{hyperref}
\usepackage{url}
\usepackage{graphicx} 
\usepackage{amsmath,amssymb,amsfonts}
\usepackage{algorithmic}
\usepackage{algorithm}
\usepackage{textcomp}
\usepackage{xcolor}
\usepackage{makecell}
\usepackage{enumitem}
\usepackage{tikz}
\usepackage{wrapfig}
\usepackage{caption}
\usetikzlibrary{positioning, arrows.meta,fit,calc}

\title{Learning Approximate Isometric Embeddings for Efficient Template Placement}

\iclrfinalcopy 
\begin{document}

\author{Alexandra Wernersson \\
Nikhef, National Institute for Nuclear Physics and High-Energy Physics \\
Amsterdam, Netherlands\\
CAISR Health, Halmstad University \\
Halmstad, Sweden \\
\texttt{alewer97@gmail.com} \\
\And
Jessica Irwin \& Melissa Lopez \\
Institute for Gravitational and Subatomic Physics (GRASP), Utrecht University \\
Nikhef, National Institute for Nuclear Physics and High-Energy Physics \\
Netherlands \\
\texttt{\{j.irwin, m.lopez\}@uu.nl} \\
\And
Sarah Caudill \\
Department of Physics and the Center for Scientific Computing and \\ Data Science Research, University of Massachusetts Dartmouth \\
Dartmouth, USA \\
\texttt{scaudill@umassd.edu} \\
}

\maketitle

\begin{abstract}
The sphere covering problem has broad applications in science however, exact solutions are computationally infeasible on curved, high-dimensional manifolds because distances must be evaluated using a position-dependent metric. We present $\texttt{isobank}$, a normalizing flow that learns an approximately isometric coordinate transformation, such that Euclidean distances in the latent space approximate metric-induced proper distances in the physical parameter space. This enables template placement using Euclidean distance checks in the latent space, avoiding repeated metric evaluations during placement.

We evaluate $\texttt{isobank}$ in the context of gravitational wave template bank construction, where the sphere covering problem corresponds to the template placement problem. On a five-dimensional eccentric binary parameter space, $\texttt{isobank}$ reaches 97\% coverage at a maximum mismatch of $\mu=0.03$ with 23\% fewer templates than the state-of-the-art stochastic placement code $\texttt{mbank}$, which reaches 94\% coverage. Coverage is measured on independent injections drawn uniformly over the masses, spins and eccentricity, using the exact waveform match rather than the metric approximation. Training the flow takes about 2.5h and placing the bank under 1h on 32 CPU cores. $\texttt{isobank}$ is publicly available on Github\footnote{\url{https://github.com/Alexandra-Wernersson/isobank}}.
\end{abstract}

\section{Introduction}

Covering a manifold $\mathcal{B} \subset \mathbb{R}^D$ with a finite set of balls is a fundamental problem with broad applications across astronomy, engineering, and medicine \citep{5ef98658-c62c-3961-914e-a3e22c4c8128}. The objective is to cover a given manifold $\mathcal{B}$ with a finite set of metric balls $b(\lambda_i) = \left\{ \lambda \in \mathcal{B}: d_g(\lambda,\lambda_i) \leq r \right\}$ with radius $r$ and distance metric $d_{g}$, such that every point $\lambda \in \mathcal{B}$  falls within at least one ball, subject to a maximum quantity of interest, \textit{e.g.}, a measure of error or cost.
The coverage or discretization of $\mathcal{B}$ naturally introduces a certain amount of error, necessitating a trade-off between the number of balls and computational cost: while a small number of metric balls $b(\lambda_i) $ will not cover  $\mathcal{B}$ sufficiently to satisfy the maximum quantity of interest, a large number of $b(\lambda_i) $ will increase the computational cost, leading to the ``curse of dimensionality" as the problem's dimension increases. Thus, the optimal solution will be composed of the smallest number of metric balls $b(\lambda_i) $  subject to a maximum quantity of interest.
In this work, we seek an approximately optimal solution to the sphere covering problem in the context of gravitational wave (GW) template bank construction, which is commonly known as a \textit{template placement} problem.

Modeled GW searches use the matched filtering method to cross-correlate detector data with a set of waveform models, known as \textit{templates}, to detect expected GW signals. These templates can be generated using analytical and numerical techniques \citep{Schmidt:2020ekt} and are parametrized by a continuous, high-dimensional parameter vector $\lambda \in \Lambda$ describing the physical properties of the GW source, such that the waveform manifold is defined as $\mathcal{B}:= \{ h(t; \lambda) \mid \lambda \in \Lambda \}$. Thus, the $i$th template is defined by its centroid $\lambda_{i}$, and the associated metric ball $b(\lambda_{i}) $ covers a region in  $\mathcal{B}$. In this context, the ``curse of dimensionality '' arises from adding physical effects to maximize GW signal discovery, which increases the size of the template banks by many orders of magnitude.
Moreover, finding the optimal solution is complex: the geometry of the waveform manifold $\mathcal{B}$ can be uneven, requiring more expensive distance evaluations and creating non-uniform regions that need different numbers of templates for adequate coverage.

To overcome these challenges, we present \texttt{isobank}: a template placement approach powered by normalizing flows. The normalizing flow learns a coordinate transformation of the waveform manifold $\mathcal{B}$ into a Euclidean latent space. In this space, the Euclidean distances approximate the proper distances\footnote{Here, the proper distance refers to the geodesic distance induced by the positive definite waveform metric $g$, obtained by minimizing $\int_\gamma \sqrt{d\theta^\mathsf{T}g(\theta)d\theta}$ over curves $\gamma$ connecting two points in $\mathcal{B}$.} induced by the waveform metric at the length scales relevant to template placement. The learned map approximately flattens the metric of $\mathcal{B}$ across $\lambda$, reducing template placement to stochastic sampling with Euclidean distance checks in the latent space. This eliminates metric evaluations during placement and substantially reduces the computational cost of constructing the template bank.

\section{Background}\label{sec:background}

Given a time series $s(t) = h(t; \lambda) + n(t)$, where $h(t; \lambda)$ is the unknown GW signal and $n(t)$ is the detector noise, the \textit{matched filtering} method can ``filter out '' the detector noise $n(t)$ via cross-correlation with $s(t)$ and the optimal filter $K(t)$. This task is challenging because of the complexity of $n(t)$, but we can simplify the problem by approximating {$n(t)$} as locally Gaussian and stationary. Thus, it can be shown that the optimal filter $K(t)$ is the {template} of the target GW signal $h_i \equiv h(t; \lambda_i)$ \citep{Owen:1998dk, Allen:2005fk}.

The morphology of the template is defined, at least, by a set of fifteen $\lambda = (\phi, \theta)$ parameters. These parameters can be divided into \textit{extrinsic} parameters $\phi$, which depend on the orientation of the astronomical event with respect to an outside observer, and \textit{intrinsic} parameters $\theta$, which are the physical characteristics of the astronomical event that are independent of the observer's location. The geometric properties of the waveform manifold $\mathcal{B}$ are then described by its metric, which is naturally introduced by the matched filtering method. The similarity between two templates $h_{i}$ and $h_{j}$ is quantified in terms of the overlap $\mathcal{O}$ defined as,

\begin{equation}
\mathcal{O}(h_{i}, h_{j}) = (h_{i} |h_{j}) := 4 \text{Re} \int^{f_{\text{upper}}}_{f_{\text{lower}}} df \frac{\tilde{h}_{i}(f) \tilde{h}_{j}^{\ast}(f)}{S_{n}(f)},
\end{equation}

{where $\tilde{\cdot}$ is the Fourier transform, $^{*}$ is the complex conjugate, the integral extends in $[f_{\text{lower}},f_{\text{upper}}]\,$Hz related to the detector's sensitivity, and  $S_{n}(f)$ denotes the one-sided power spectral density of detector noise. }
{It is customary to normalize the templates, so that $\mathcal{O \in [\text{0, 1}]}$, \textit{i.e.} two identical templates yield $\mathcal{O}=1$, while two orthogonal templates yield $\mathcal{O}=0$. Given $h_{i}$ and $h_{j}$ parametrized by $\lambda_{i}$ and $\lambda_j = \lambda_{i} + \Delta \lambda$, respectively, we can define the match $\mathcal{M}$ between them as the maximized overlap with respect to the extrinsic parameters $\phi$ such that}
$\mathcal{M}(\theta_{i}, \theta_{i} + \Delta \theta) =
\max_{\phi, \Delta \phi} \mathcal{O}(\theta_{i}, \phi_{i}, \theta_{i} + \Delta \theta, \phi_{i} + \Delta \phi)$.
{In general, $\mathcal{M}$ is not symmetric and does not satisfy the triangular inequality, but it introduces a sense of ``closeness" between two templates dependent on $\Delta \theta$. Trivially, we can define the \textit{mismatch} as }
$
\mu(\theta_{i}, \theta_{i} + \Delta \theta) = 1 - \mathcal{M}(\theta_{i}, \theta_{i} + \Delta \theta),
$
{ which is the \textit{quantity of interest} to limit when solving this problem. Now, we approximate the match $\mathcal{M}$ as a second-order Taylor expansion around $
\Delta \theta = 0$. }

\begin{equation}
\begin{aligned}
\mathcal{M}(\theta, \theta + \Delta \theta) \approx 1 + \frac{1}{2}\frac{\partial^2 \mathcal{M}}{\partial \Delta \theta^{i} \partial \Delta \theta^{j}}\Delta \theta^{i}\Delta \theta^{j} \Longrightarrow g_{ij}(\theta) := - \frac{1}{2}\left.
\frac{\partial^2 \mathcal{M}}
{\partial \Delta\theta^i \partial \Delta\theta^j}
\right|_{\Delta\theta=0}
\end{aligned}
\end{equation}

where $g(\theta) \equiv g_{ij}(\theta)$ is defined as the metric of the waveform manifold $\mathcal{B}$ over the intrinsic parameters $\theta$.
To evaluate the coverage of the template bank, it is customary to conduct a \textit{simulated injection\footnote{In GW data analysis, an injection is the result of adding a GW signal to detector noise.} campaign}: $h(\theta')$ waveforms are generated by sampling $\theta'$ source parameters from a prior distribution, and added to simulated detector noise\footnote{To generate the detector noise, we use the average power spectral density of LIGO Hanford during the third observing run (O3); \texttt{aligo\_O3actual\_H1.txt}.}. To quantify the coverage, we compute the \textit{fitting factor} (FF) for each source parameter $\theta'$ as,

\begin{equation}\label{eq:FF}
    \mathrm{FF}(\theta') =
    \max_{\theta \in \mathcal{B}} \mathcal{M}(\theta', \theta).
\end{equation}

For a chosen maximum mismatch $\mu$, an injection is considered \textit{recovered} if its fitting factor satisfies \begin{equation} \mathrm{FF}(\theta') \geq 1-\mu, \end{equation} and \textit{missed} otherwise. We define the \textit{coverage} of the template bank as the fraction of injections that are recovered.

\section{Related work}
A breadth of methods have been implemented in GW to solve the sphere covering problem, also known as the template placement problem. We can categorize them into four broad approaches:

\begin{itemize}
\item[-] \textit{Geometric:} {it solves the \textit{sphere covering problem} in the waveform manifold $\mathcal{B}$. Here,  $g(\theta)$ is defined analytically using a second-order Taylor expansion, defining a regular lattice for placing templates in a locally flat space. This method struggles with waveform manifolds $\mathcal{B}$ that have intrinsic curvature, complex boundaries, or high dimensionality \citep{ Harry:2013tca, PhysRevD.99.123022, Phukon_2025, Cabass:2025wqg}.}
\item[-] \textit{Stochastic:} {new templates are drawn from the parameter space $\Lambda$ of interest, and are added to the template bank when they sufficiently differ from existing ones. This method evaluates the distance between templates; while it can be extended to arbitrary parameter spaces $\Lambda$, the cost per evaluation grows steeply with dimensionality and produces redundant template banks \citep{Harry_2009, Ajith:2012mn, Kacanja:2024pjh}.}
\item[-] \textit{Hybrid:} {it employs geometric placement in regions of the parameter space $\Lambda$ where it is most efficient, and stochastic placement where the latter provides improved coverage \citep{PhysRevD.95.104045, Roy:2017oul, PhysRevD.109.124049}.}
\item[-] \textit{Random:} { new templates are drawn from a probability distribution proportional to the local metric volume element of the waveform manifold $\mathcal{B}$, reducing the computational cost of bank construction. Although random placement generally produces larger template banks than other approaches, it scales favorably to high-dimensional parameter spaces \citep{PhysRevD.105.102003, Coogan:2022qxs, Schmidt:2023gzj}.}
\end{itemize}

The most widely used method in GW data analysis is stochastic template placement. However, this method is prohibitive for the large banks that high-dimensional spaces demand. \texttt{isobank} falls within the \textit{geometric template placement} category, as the trained flow learns the coordinate transformation from the waveform manifold $\mathcal{B}$ to an approximately isometric space. In this way, the trained flow removes both obstructions at once: distances become Euclidean in the latent space, so each acceptance test reduces to a nearest-neighbor query in a $k$-d tree, and \emph{no waveform or metric evaluations are performed during placement}. 
While similar work uses autoencoders \citep{Cabass:2025wqg}, \texttt{isobank} leverages the mathematical properties of normalizing flows for efficient template placement in higher-dimensional spaces.
We compare \texttt{isobank} against \texttt{mbank} \citep{Schmidt:2023gzj}, which also uses normalizing flows but for metric volume sampling rather than learning a distance preserving representation. While \texttt{mbank} uses the flow to sample regions of high metric volume, \texttt{isobank} learns coordinates in which proper distances are approximated by Euclidean distances, enabling efficient nearest-neighbour template placement.

\section{Methodology}
\subsection{Normalizing flows}

Normalizing flows learn invertible transformations between coordinate spaces, typically parameterized by neural networks \citep{Kobyzev:2019ydm}. Consider a parameter vector $\theta\in\Lambda$, which parametrizes the waveform manifold $\mathcal{B}$, and a latent coordinate vector $z\in\mathbb{R}^D$ related by a bijective transformation
$z=f_{\psi}(\theta),$ with Jacobian $J=\partial f_{\psi}/\partial\theta$. The change-of-variables formula relates probability densities 
according to
$
p_\theta(\theta)
=
p_z(f_{\psi}(\theta))
\left|\det J_{\psi}(\theta)\right|.
$
In conventional applications, this relation represents a complex probability distribution through a simple latent distribution and enables efficient sampling via the inverse transformation. Here, we instead exploit the invertibility and flexible nonlinear structure of normalizing flows to learn a coordinate transformation between the physical and latent spaces.

We adopt a RealNVP-based architecture~\citep{dinh2017realnvp} featuring coupling layers that split the input in two: $z_a = \theta_a$, while $z_b = \theta_b\odot\exp[s_{\psi}(\theta_a)] + t_{\psi}(\theta_a),$ and $\theta_a$ parametrizes an affine transformation. Note that $s_{\psi}$ and $t_{\psi}$ are neural networks that determine the scaling and translation of $\theta_b$, and $\odot$ denotes element-wise multiplication. The transformation is invertible, and its Jacobian is triangular and therefore has a tractable determinant. By cycling the masking pattern between coupling layers, every coordinate is both transformed and used as a conditioning variable.

\subsection{Learning an approximately isometric embedding}

This work aims to bypass repeated metric evaluations during template placement in highly curved parameter spaces. Here, the flow does not model a probability distribution or generate samples. Instead, the learned forward transformation $f_{\psi}$ provides a coordinate map,
\begin{equation}
\mathcal{B} \subset \mathbb{R}^D
\xrightarrow{\quad f_{\psi}\quad}
z\in\mathbb{R}^D,
\end{equation}
chosen such that Euclidean distances in the latent space approximate proper distances in the physical parameter space.    
For Euclidean latent distances to reproduce proper distances, the flat Euclidean line element must pull back to $g(\theta)$. Writing the differential as $dz = J_{\psi} d\theta$ requires
$
dz^\top dz = d\theta^\top (J_{\psi}^\top J_{\psi})\, d\theta = d\theta^\top g(\theta)\, d\theta,
$
which holds for arbitrary displacements iif the network's Jacobian satisfies:
\begin{equation}
    J_{\psi}^\top J_{\psi} = g(\theta).
    \label{eq:isometry}
\end{equation}

Although this relation can always be satisfied locally at a single point, it cannot in general be realized by a single coordinate transformation of the same dimension over a curved region. Exact isometric embeddings can instead require a higher-dimensional Euclidean space \citep{nash1956}.

Consequently, we settle for an approximate, scale-targeted embedding. We specifically optimize for distance preservation at the finite scales relevant for template placement. For two points $\theta_1$ and $\theta_2$, we define distance in the latent space as the Euclidean norm of their difference, $d_z(\theta_1,\theta_2) = \bigl\| f_{\psi}(\theta_1) - f_{\psi}(\theta_2) \bigr\|_2$, and compare it with their proper distance $d_{\mathrm{proper}}$ under the physical metric. Exact distance preservation at a given scale corresponds to $d_z / d_{\mathrm{proper}} = 1$.

\subsection{Multi scale distance-matching loss}
\label{sec:distance-matching loss}

{To achieve the desired performance, we need to enforce distance preservation at different scales. For this aim, the} training objective is a weighted sum of four distance-matching terms\footnote{$\beta_{1} =5, \beta_{2} =10, \beta_{3} =10$ are coefficients obtained empirically to balance the loss function.},
distinguished by the distribution from which training pairs are drawn, as summarized in Table \ref{tab:pairs}.
\begin{equation}
    \mathcal{L} =
      \mathcal{L}_{\mathrm{glob}}
      + \beta_{1}\,\mathcal{L}_{\mathrm{loc}}
      + \beta_{2}\,\mathcal{L}_{\mathrm{bnd}}
      + \beta_{3}\,\mathcal{L}_{\mathrm{edge}},
    \label{eq:total-loss}
\end{equation}

\paragraph{Global term $(\mathcal{L}_{\mathrm{glob}})$:} Pairs $(\theta_1, \theta_2)$ are drawn independently
from the full parameter space, and we penalize the squared difference of log
distances,
$
    \mathcal{L}_{\mathrm{glob}}
      = \mathbb{E}\Bigl[ \bigl( \log d_z - \log d_{\mathrm{proper}} \bigr)^2 \Bigr] ,
$
which is scale-invariant and therefore weights the (large) global distances on an equal footing with local ones. This term preserves the global ordering of distances and prevents the flow from collapsing distant regions of the manifold without competing with the local terms for accuracy at any particular scale.

\paragraph{Calibration:}
Before training, we determine the metric distance corresponding to the
quantity of interest: the maximum mismatch $\mu$. For sufficiently small parameter displacements, the waveform mismatch is related to the quadratic metric by
$
    1-\mathcal{M}(\theta,\theta+\Delta\theta)
    \simeq
    [\Delta\theta^{\mathsf T}
    g(\theta)\Delta\theta]/2
    =
    d_{\mathrm{proper}}^2/2.
$
Thus, under the infinitesimal quadratic approximation, a target mismatch of $\mu=0.03$ corresponds to a proper distance
\begin{equation}
d_{\mathrm{proper}} \simeq \sqrt{2\mu} \simeq 0.245.
\end{equation}
At the finite separations relevant for template placement, however, the quadratic metric approximation does not accurately reproduce the true waveform mismatch. We therefore empirically calibrate the target placement distance by generating metric-guided pairs over a range of separations and evaluating their true waveform matches. We define $r_{\mathrm{target}}$ as the proper distance at which the mean true waveform mismatch reaches $\mu=0.03$.

\paragraph{Metric-guided local pairs:}
The remaining terms use pairs constructed at controlled separations relevant for template placement. Given an anchor point $\theta_1$, we draw a direction $v$ uniformly from the Euclidean unit sphere, as $v \sim \mathcal{U}(S^{d-1})$, and a dimensionless scale factor $\alpha$. Since the physical metric is generally anisotropic, the same coordinate displacement corresponds to different metric distances depending on its direction. We therefore normalize $v$ by its local metric norm $|v|_{g}$ to ensure consistent physical spacing, generating the second point as,
\begin{equation}\label{eq:metric-guide}
    \theta_2 =
    \theta_1
    + \alpha r_{\mathrm{target}}
    \frac{v}{|v|_{g}}
    =
    \theta_1
    + \alpha r_{\mathrm{target}}
    \frac{v}{\sqrt{v^\top g(\theta_1)v}}.
\end{equation}
This construction ensures that
$
\Delta\theta^\top g(\theta_1)\Delta\theta
=
(\alpha r_{\mathrm{target}})^2,
$
so the local quadratic metric assigns the displacement the same length, $\alpha r_{\mathrm{target}}$, regardless of direction $v$. The resulting pair is therefore generated at approximately the desired metric scale; we then compute the actual proper distance between the two points from the metric and use it as the training target. 

The metric-guided local pairs use a relative (ratio) loss with \textit{adaptive per-pair weights} $\gamma_i$,
\begin{equation}
   \mathcal{L}_{\bullet}
     = \frac{1}{N} \sum_{i=1}^{N} \gamma_i \, (\frac{d_{z,i}}{d_{\mathrm{proper},i}} - 1)^2 .
   \label{eq:ratio-loss}
\end{equation}

Since not all sampled pairs contribute equally to the final coverage quality, we reweight the loss by the current embedding error: at every gradient step, the worst fitting fraction of the batch, ranked by ${|d_{z,i} - d_{\mathrm{proper},i}|}/{d_{\mathrm{proper},i}}$, receives an enhanced weight while the rest keep unit weight.
Early in training the errors are large and roughly
uniform, so the weighting is nearly inactive; as the fit converges it
concentrates effort on the genuinely hard pairs, predominantly those near the boundary of the parameter space. Moreover,   normalizing Eq.~\ref{eq:ratio-loss} by $d_{\mathrm{proper}}$ places all sampled
scales on an equal footing; an unnormalized loss
$(d_z - d_{\mathrm{proper}})^2$ would be dominated by the largest separations. {In the following, we describe the different metric-guided terms of Eq.~\ref{eq:ratio-loss}:}
\begin{enumerate}[label=(\roman*)]
\item \textit{Local term $(\mathcal{L}_{\mathrm{loc}})$:} $\alpha \sim \mathcal{U}[0.1,,2.5]$, spanning separations from $0.1,r_{\rm target}$ to $2.5,r_{\rm target}$; $\alpha = 1$ corresponds to the covering radius itself.

\item \textit{Boundary term $(\mathcal{L}_{\mathrm{bnd}})$:} Identical to the $\mathcal{L}_{\mathrm{loc}}$ but with
$\alpha \sim \mathcal{U}[0.6,\, 1.4]$, densely sampling the window of
separations that directly determines whether a point is covered by its nearest
template. Its weight in Eq.~\ref{eq:total-loss} is twice that of the local
term, reflecting its direct impact on coverage.

\item \textit{Edge term $(\mathcal{L}_{\mathrm{edge}})$:} Identical to $\mathcal{L}_{\mathrm{bnd}}$, but with anchors
$\theta_1$ drawn from a hard subregion that is underrepresented with the uniform anchor sampling. Note that for our \textit{toy example}, the $\mathcal{L}_{edge}$ is not used since the difficulty lies in the global anisotropy covered by the $\mathcal{L}_{loc}$ and $\mathcal{L}_{bnd}$. For our \textit{high-dimensional example}, we introduce the $\mathcal{L}_{edge}$ term and sample the high-eccentricity subregion $e \geq 0.12$, where the
local geometry is most degenerate. Without this term, sparse edge regions dominate the resulting bank's coverage failures.
\end{enumerate}

\subsection{Training strategy}
\label{sec:pipeline}

The complete training workflow consists of calibrating the target placement radius, generating metric-guided training pairs, optimizing the flow, and selecting the model. {In the following, we provide details on this workflow for reproducibility.}

\begin{table}[]
\centering
\caption{Number of pairs computed for a given scale $\alpha$, and  anchor point $\theta_1$ for each loss term. }
\begin{tabular}{cccc}
\hline
\textit{Term}                         & \textit{Number of Pairs} & \textit{$\alpha$ scale} & $\theta_2$ constraints     \\ \hline
Global $\mathcal{L}_{\mathrm{glob}}$  & $10^4$                   & -                       & Full parameter space       \\
Local $\mathcal{L}_{\mathrm{loc}}$    & $5\times10^3$            & $[0.1, 2.5]$            & Full parameter space       \\
Boundary $\mathcal{L}_{\mathrm{bnd}}$ & $10^4$                   & $[0.6, 1.4]$            & Full parameter space       \\
Edge $\mathcal{L}_{\mathrm{edge}}$    & $10^4$                   & $[0.6, 1.4]$            & Eccentricity $e \geq 0.12$ \\ \hline
\end{tabular}
\label{tab:pairs}
\end{table}

 \paragraph{Pair generation:}
 We generate all training pairs before optimization, so waveform dependent computations are confined to preprocessing. We summarize in Table \ref{tab:pairs} the number of pairs for each term of the training objective defined in Eq. \ref{eq:total-loss}. For the global term, pairs are drawn independently from the full parameter space; since such sampling produces few pairs at placement relevant separations, the metric-guided local pairs are instead generated using Eq.~\ref{eq:metric-guide}. For each pair, we evaluate the proper distance under the metric once and store it for subsequent training.

\paragraph{Optimization and model description:} 
We train the flow for $10^3$ epochs using Adam~\citep{Kingma:2014vow} with an initial learning rate of $5\times10^{-4}$, cosine annealing~\citep{Loshchilov:2016tgk} to $10^{-6}$, and gradient norm clipping. The model minimizes the weighted sum in Eq.~\ref{eq:total-loss}.
We use a RealNVP-type flow~\citep{dinh2017realnvp} with nine affine coupling layers whose binary masks cycle through three patterns arranged so that every coordinate is both transformed and used as a conditioning variable within each three-layer cycle. Each coupling transformation is parametrized by a single hidden layer network with $256$ neurons and a $\tanh$ activation function. The flow required approximately 2.5 hours to train on 32 CPU cores, and template placement less than one hour.

\paragraph{Template Placement Algorithm:}
With the trained flow, we perform template placement using Euclidean distances in the latent space, as summarized in Algorithm~\ref{algorithm:templates} in Sec.~\ref{sec:appendix_methods}. We initialise an empty template bank $\hat{\mathcal{B}}$ and repeatedly draw batches of candidate points ${\theta_i}$ from a proposal distribution $q(\theta)$ over the physical parameter space. Each candidate is mapped to latent space, $z_i=f_\psi(\theta_i)$, and accepted if its nearest existing template is farther than the target placement radius,

$$
\min_{z_j\in\hat{\mathcal{B}}}\|z_i-z_j\| > r_{\mathrm{target}}.
$$

Nearest-neighbour queries are performed using a $k$-d tree constructed from the latent coordinates of the current bank, avoiding explicit pairwise comparisons with all existing templates.

Placement is terminated when the estimated fraction of uncovered metric volume falls below a prescribed tolerance $\varepsilon$. The volume estimate uses the Jacobian determinant of the flow as an approximate metric-volume weight, with importance weighting to account for the proposal distribution. This criterion is only a stopping condition; coverage is assessed independently with true waveform matches. In practice we operate at a coarser effective radius than $r_{\rm target}$, trading bank size against coverage (Appendix~\ref{sec:appendix_methods}).

\section{Data set}

\begin{table}[]
\centering
\caption{Description of the parameter spaces.}
\begin{tabular}{lcc}
\hline
\textit{Space} & \textit{Parameters} & \textit{$D$} \\ \hline
\textit{Toy example:} aligned-spin & \makecell[l]{$M \in [30, 50]\,M_\odot$ \\
               $q \in [1, 5]$ \\
               $\chi \in [-0.99, 0.99]$ \\
               \texttt{IMRPhenomD}~\citep{imrphenomd}} & 3 \\ \hline
\textit{High-dimensional example:} eccentric & \makecell[l]{$m_1 \in [1, 7]\,M_\odot$ \\
             $m_2 \in [1, 3]\,M_\odot$ \\
             $s_{1z} \in [-0.9, 0.9]$ \\
             $s_{2z} \in [-0.05, 0.05]$ \\
             $e \in [10^{-5}, 0.15]$ \\
             \texttt{TaylorF2Ecc}~\citep{taylorf2ecc}} & 5 \\ \hline
\end{tabular}
\label{table:param_spaces}
\end{table}

\begin{figure}[]
    \centering
    \begin{minipage}[c]{0.60\columnwidth}
        \centering
        \includegraphics[width=\linewidth]{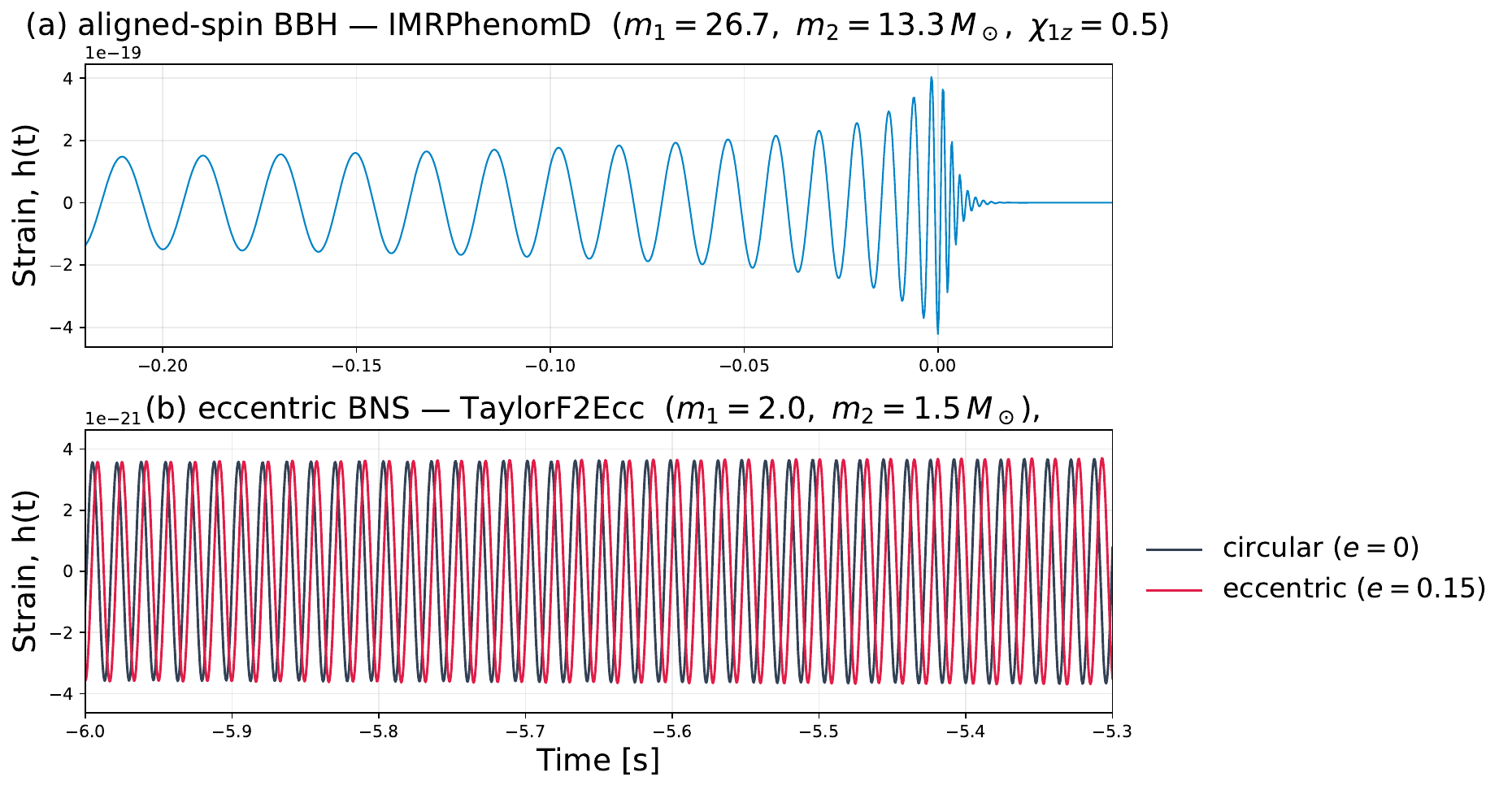}
    \end{minipage}
    \hfill
    \begin{minipage}[c]{0.39\columnwidth}
        \captionof{figure}{%
        \textit{(Top panel)} The aligned-spin case with the \texttt{IMRPhenomD} waveform model showing inspiral, merger and ringdown phase. \textit{(Bottom panel)} Eccentric case with the \texttt{TaylorF2Ecc} inspiral waveform. Demonstrating the case for an inspiraling binary neutron star (BNS) signal with $e=0$ and $e=0.15$ to show that eccentric imprint is a slow phase drift rather than a merger feature.
        }
        \label{figure:waveform_models}
    \end{minipage}
\end{figure}

{To test the effectiveness of our approach, we initially apply it to a \textit{toy example} to learn a three-dimensional \textit{aligned-spin} space, representative of binary black hole (BBH) systems, parameterized by the total mass $M$, mass ratio $q$, and spin parameter $\chi$, and modeled by \texttt{IMRPhenomD}~\citep{imrphenomd},  as shown in Table~\ref{table:param_spaces}. 

We then apply our method to a more challenging five-dimensional \textit{eccentric aligned-spin} parameter space covering both binary neutron star (BNS) and neutron star-black hole (NSBH) systems. The parameter space is defined by the component masses $m_1$ and $m_2$, the aligned spin projections $s_{1z}$ and $s_{2z}$\footnote{Spins are defined along the orbital angular momentum ($\hat{z}$) axis. The \textit{toy example} parameterizes spin as a 1D scalar $\chi$ (a weighted linear combination of binary spins), while the \textit{high-dimensional} example retains the full 2D component spin parameterization with $s_{1z}$ and $s_{2z}$.}, and the eccentricity $e$, using the model \texttt{TaylorF2Ecc}~\citep{taylorf2ecc}. Eccentricity adds variety and structure to the waveform parameter and a strongly curved waveform manifold metric.

BNS are expected to undergo significant orbital circularization before entering the sensitive band of GW detectors, so they have low eccentricities by the time they become detectable. Following \cite{Phukon_2025}, we choose $e_{\mathrm{max}}=0.15$ as a conservative upper bound, encompassing systems with significant residual eccentricity while remaining within the moderate eccentricity regime considered relevant for low-mass compact binaries. The computational challenge associated with eccentricity is illustrated in \cite{Lenon_2021}, which demonstrated the feasibility of constructing stochastic template banks for eccentric binaries for Cosmic Explorer. They found that increasing the maximum eccentricity from $e\leq0.05$ to $e\leq0.1$ approximately doubled the bank size to 4.5 million templates.

\section{Results}
\subsection{Distance preservation validation}
We evaluate several diagnostics of the assumption that the trained flow has
learned an approximate isometry. The most direct one is the comparison of physical proper distances against latent distances for pairs of points not used in training. For a perfect isometry, all pairs should lie on the diagonal represented as a black dashed line in Fig.~\ref{figure:distance_preservation_shells}.
The physical coordinates carry almost no distance information: even after optimal global rescaling, coordinate distances are essentially uncorrelated with the proper distance and would therefore appear as a scattered cloud rather than along the diagonal. On the left panel of Fig.~\ref{figure:distance_preservation_shells}, we show that the same pairs collapse onto the diagonal after the flow has learned the high dimensional problem of the eccentric space. Note that the vertical red line marks the calibrated $r_{\mathrm{target}}$, below which the scale is relevant for template placement.

\begin{figure}[h]
\centering
\begin{minipage}[c]{0.48\columnwidth}
\centering
\includegraphics[width=\linewidth]{
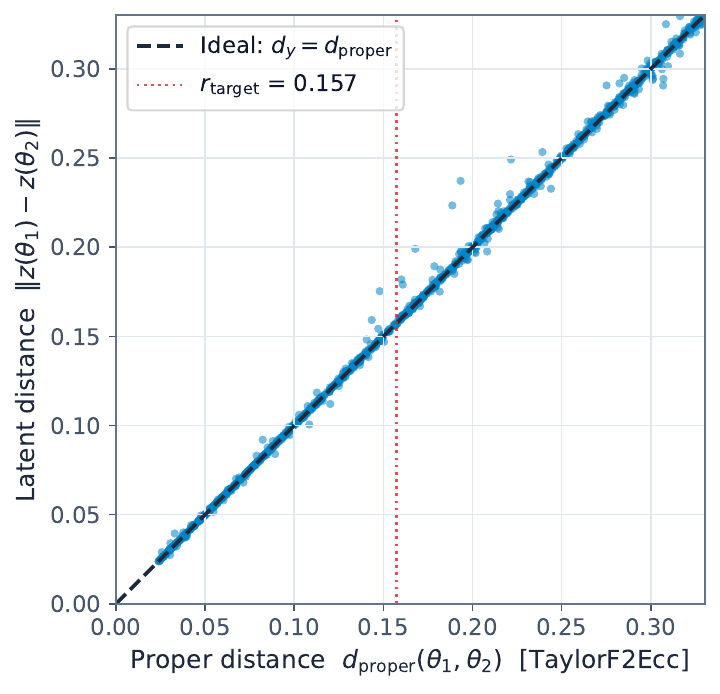
}
\end{minipage}
\hfill
\begin{minipage}[c]{0.48\columnwidth}
\centering
\includegraphics[width=\linewidth]{
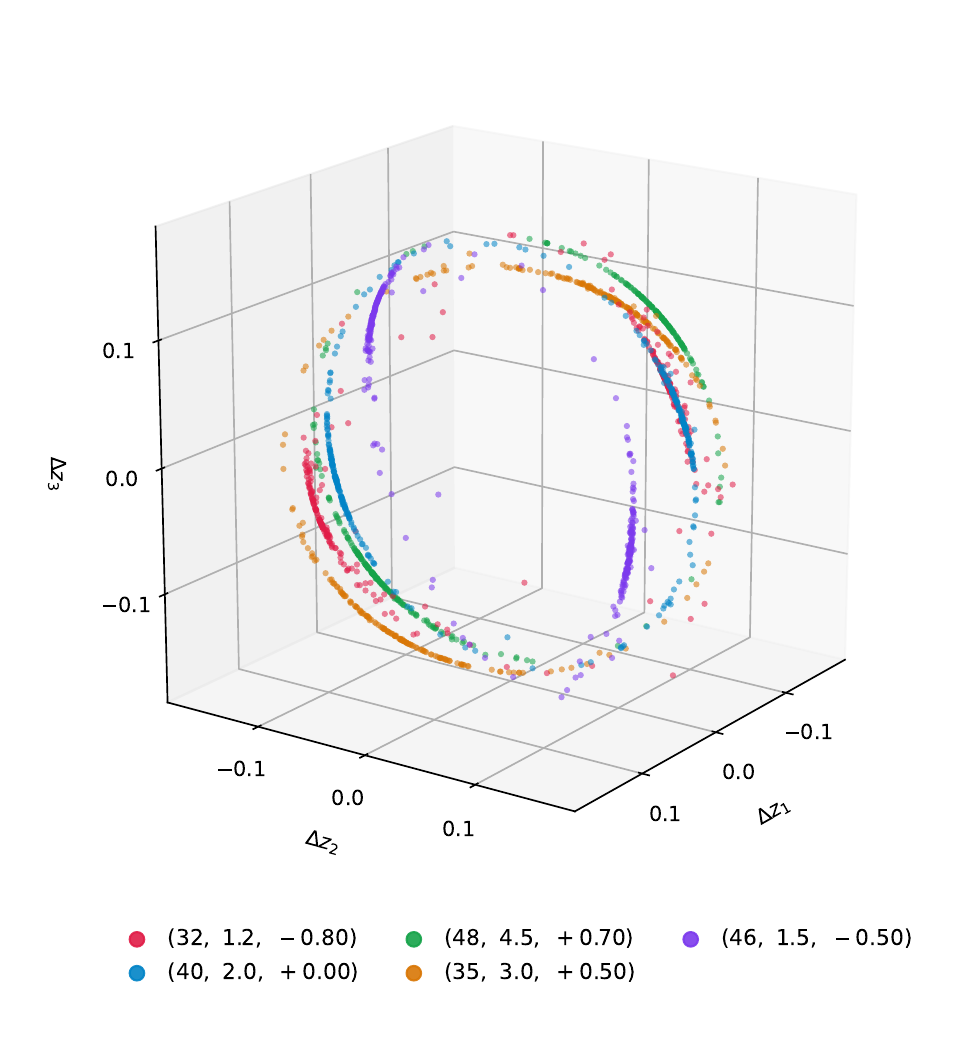
}
\end{minipage}
\caption{%
\emph{Left:} Distance preservation evaluated on the eccentric 5D space on test pairs generated independently of the training set, showing proper distance against the Euclidean distance in the flow's latent space. The black dashed diagonal marks exact isometry.
\emph{Right:} Transformation of fixed proper distance mismatch shells into the flow's latent space for the 3D aligned-spin parameter space. Points lying at a fixed proper distance $r_{\rm target}$ from each anchor point become approximately spherical, illustrating the approximate isometry of the learned representation.
}
\label{figure:distance_preservation_shells}
\end{figure}

\subsection{Flow transformation}

Another test of our method's feasibility is to visualize how the flow transforms the local geometry of the parameter space. Around a set of
anchor points, we construct \emph{unit mismatch shells}, points lying at a fixed proper distance $r_{\rm target}$ from each anchor point, and map them through
the flow. For ease of visualization, we show this in Fig.~\ref{figure:distance_preservation_shells} only for the 3D aligned-spin space, whose three dimensions can be displayed directly. The same construction extends to the 5D eccentric space. 
We translate each shell so that its anchor lies at a common origin. In the physical coordinates, equal proper distance corresponds to highly anisotropic coordinate displacements, producing elongated shells. Under the learned transformation, these same points are mapped to approximately spherical shells with radii close to $r_{\rm target}$, as shown in the right panel. Thus, the flow transforms the strongly anisotropic local geometry induced by the waveform metric into approximately Euclidean geometry, such that proper distances relevant for template placement are represented by ordinary Euclidean distances in the latent space. This approximate isometry allows template placement and nearest-template searches to use standard Euclidean operations without evaluating the metric.

\subsection{Effectualness of the template bank}

{In Fig.~\ref{figure:FF}, we compare the performance of \texttt{isobank} with respect to \texttt{mbank} \citep{Schmidt:2023gzj} for the aligned-spin and eccentric spaces using the O3 PSD, with $10^4$ injections in each space. Each curve shows the cumulative fraction of injections with a fitting factor below a given value. It is standard to set a threshold at $\text{FF} = 0.97$ \citep{Owen:1998dk}, so the height of each curve at this threshold is the fraction of missed signals. A curve that stays low until close to $\text{FF} = 1$ indicates good coverage, whereas a long tail towards low FF reveals regions where the bank matches signals poorly.}

In the aligned-spin example, both codes perform equally well: \texttt{mbank} covers $99.80 \pm 0.04\%$ of the injections with $23{,}167$ templates, and \texttt{isobank} covers $99.72 \pm 0.05\%$ with $23{,}810$ templates. In the challenging eccentric example, the \texttt{mbank} bank of $2.95$M templates, generated with its default settings at $\mu = 0.03$, covers $94.2 \pm 0.2\%$ of the injections and shows a long tail below $\text{FF} = 0.97$, reaching $\text{FF} = 0.61$ for the worst injection. \texttt{isobank} covers $96.6 \pm 0.2\%$ with $2.27$M templates, $23\%$ fewer, and its worst injection has $\text{FF} = 0.93$. \texttt{isobank} therefore achieves higher coverage with fewer templates in the challening eccentric space.

As we present in Table \ref{table:FF}, \texttt{isobank} misses (does not recover) $\sim 3.36 \%$ of the joint BNS and NSBH parameter space for the eccentric example, while in \cite{Phukon_2025}, they miss $5.68\%$ of signals with 4.78M templates. However, this is not a complete comparison, as they use a different detector noise for their injection campaign: the PSD of the fourth observing run (O4). We leave such a comparison for future work. 
We also constructed an \texttt{isobank} bank for the parameter space of the O3 eccentric search of \cite{Dhurkunde:2023qoe} at the same minimal match of $0.95$, obtaining $4.46$M templates that recover $99.6\%$ of injections at $\mathrm{FF} \geq 0.95$. However, their bank also spans a source-orientation parameter that ours does not, so this comparison is indicative only and we leave an apples-to-apples comparison for future work.

\begin{figure}[]
    \centering
    \begin{minipage}[c]{0.60\columnwidth}
        \centering
        \includegraphics[width=\linewidth]{
            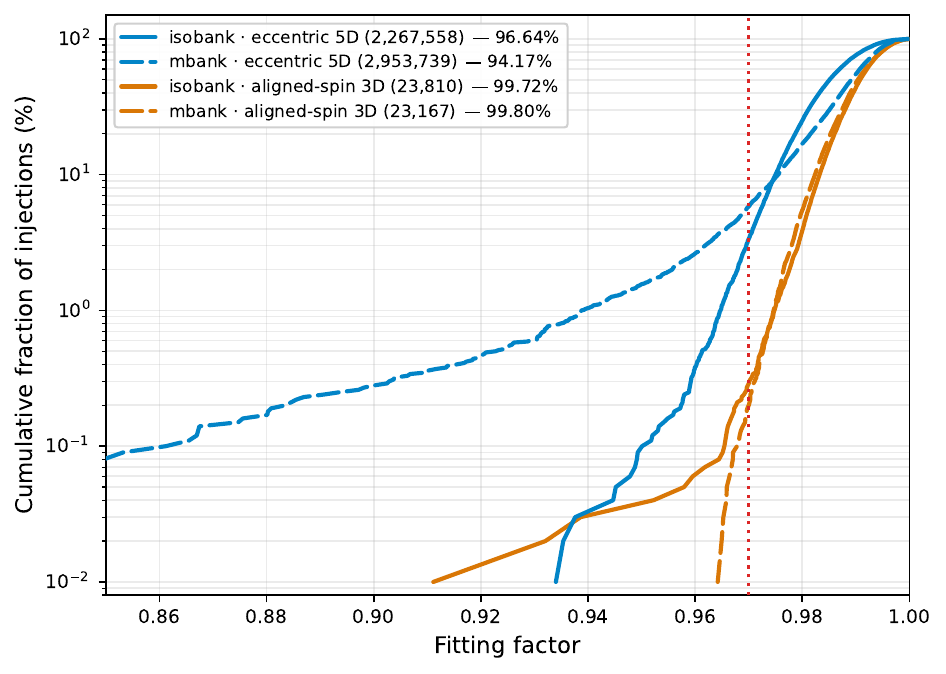
        }
    \end{minipage}
    \hfill
    \begin{minipage}[c]{0.39\columnwidth}
        \captionof{figure}{%
         Cumulative distribution of fitting factors for $10^4$ injections per parameter space, recovered with \texttt{isobank} (solid) and \texttt{mbank} (dashed) banks in the 3D aligned-spin (orange) and 5D eccentric (blue) spaces, using the O3 PSD. The red dotted line marks $\mathrm{FF} = 0.97$; the height of each curve there is the fraction of missed injections. The legend gives each bank's size (templates) and coverage ($\%$). Curves that stay low until close to $\mathrm{FF} = 1$ indicate good coverage, whereas long tails towards low FF reveal regions where the bank matches signals poorly.}
        \label{figure:FF}
    \end{minipage}
\end{figure}

\begin{table}[h!]
\caption{Percentage of the missed (not recovered) injections by \texttt{isobank} that yield FF$<0.97$ for different simulation campaigns.}
\label{table:FF}
\centering
\begin{tabular}{cc|ccc}
\multirow{2}{*}{\begin{tabular}[c]{@{}c@{}}Parameter\\ space\end{tabular}} &
\multirow{2}{*}{\begin{tabular}[c]{@{}c@{}}Bank\\ size\end{tabular}} &
\multicolumn{3}{c}{FF $<0.97$ [\%]} \\ \cline{3-5}
\rule{0pt}{2.6ex} & & BNS & NSBH & All \\ \hline
\begin{tabular}[c]{@{}c@{}}\textit{Toy example:}\\ 3D aligned-spin (BBH)\end{tabular}
  & 23,810 & -- & -- & $0.28\pm0.05$ \\
\begin{tabular}[c]{@{}c@{}}\textit{High-dimensional example:}\\ 5D eccentric\end{tabular}
  & 2.27M & $3.90\pm0.27$ & $3.80\pm0.27$ & $3.36\pm0.18$
\end{tabular}
\end{table}

\section{Conclusions}

In this work, we address the sphere covering problem in the context of GW template bank construction or template placement. For this, we develop \texttt{isobank}, whose central idea is to learn a coordinate transformation, using normalizing flows, that approximately preserves the physical geometry of the GW waveform manifold. The flow is trained using the waveform metric at multiple distance scales, producing an approximately Euclidean latent representation in which Euclidean distances approximate proper distances on the physical
manifold. Stochastic template placement can then use Euclidean nearest-neighbor queries, without repeated metric evaluations during placement. 

We demonstrate that \texttt{isobank} can efficiently construct template banks in a challenging five-dimensional eccentric parameter space. At a maximum mismatch of $\mu=0.03$, \texttt{isobank} achieves $96.6\%$ coverage while using $23\%$ fewer templates than the state-of-the-art template placement code \texttt{mbank}, which achieves $94.2\%$ coverage. These results demonstrate that learning an approximately isometric representation can enable more computationally efficient placement while achieving higher coverage with fewer templates for high-dimensional waveform manifolds.

A natural extension of this work would be to apply \texttt{isobank} to higher-dimensional waveform models incorporating additional physical effects such as precession, higher-order modes, and matter effects. The computational advantage of the learned coordinate transformation is expected to become increasingly relevant as the dimensionality and complexity of the waveform manifold increase, potentially enabling more efficient searches for signals carrying new information about their astrophysical origins.

The present construction uses a flow with the same dimensionality as the physical parameter space and therefore cannot, in general, provide a
global exact isometric map for a curved manifold. As noted in \cite{Schmidt:2023gzj}, an interesting direction would be to investigate embeddings into a higher dimensional Euclidean space, where the additional dimensions may provide greater flexibility for representing the intrinsic geometry. Such an extension could also open the possibility of studying whether the resulting representation admits useful generative or sampling procedures.

\section*{Acknowledgments}
We thank Stefano Schmidt for useful discussions in the initial phases of this project. A.W. is supported by the KKS foundation and  gratefully acknowledges the computer resources at the National Supercomputer Snellius. J. I. and M.L. is supported by the research program of the Netherlands Organisation for Scientific Research (NWO). S.C. is supported by the National Science Foundation under Grant No. PHY-2309332. 
 
\section*{AI disclosure section}
In this work, we used generative AI tools to implement methods as well as to design and provide feedback on numerical experiments. We have not used generative AI tools for translation or for qualitative and thematic data analysis, and the remaining tasks requiring disclosure are not applicable to this work. Additionally, we used generative AI tools to create and modify figures and to edit parts of the manuscript. We have reviewed all AI-assisted work: AI-generated code was verified and tested for correctness by one author. We take responsibility for the final content of this work, including text, claims or artifacts produced with the aid of generative AI.

\bibliography{iclr2026_conference}
\bibliographystyle{iclr2026_conference}

\section{Appendix}

\subsection{Methodological details}
\label{sec:appendix_methods}

In Section~\ref{sec:pipeline}, we described the template placement procedure. Here, we provide its algorithmic form and details of the uncovered volume stopping criterion. Note that $\hat{\mathcal{B}}$ represents the template bank which is an approximation of the waveform manifold $\mathcal{B}$.

\begin{algorithm}[H]
\caption{Flow-based stochastic template placement}
\label{algorithm:templates}
\begin{algorithmic}[1]

\REQUIRE trained flow $f_\psi$; parameter space $\Theta$;
placement radius $r_{\mathrm{target}}$;
proposal density $q(\theta)$ on $\Theta$;
batch size $N_b$; stopping tolerance $\varepsilon$;
maximum number of batches $N_{\max}$;
$\min_{\emptyset}\equiv+\infty$

\STATE $\hat{\mathcal{B}}\leftarrow\emptyset$
\hfill\COMMENT{template bank}

\FOR{$b=1,\ldots,N_{\max}$}

\STATE $\hat{\mathcal{B}}_0\leftarrow\hat{\mathcal{B}}$

\STATE Draw $\{\theta_i\}_{i=1}^{N_b}\sim q(\theta)$ and set
$z_i=f_\psi(\theta_i)$; build a KD-tree from
$\{f_\psi(\theta):\theta\in\hat{\mathcal{B}}_0\}$

\STATE $\mathcal{A}\leftarrow\emptyset$
    \hfill\COMMENT{templates accepted in this batch}

\FOR{$\theta_i$}

    \IF{$\displaystyle
    \min_{z_j\in\hat{\mathcal{B}}_0\cup\mathcal{A}}
    \|z_i-z_j\|>r_{\mathrm{target}}$}

        \STATE $\mathcal{A}\leftarrow
        \mathcal{A}\cup\{\theta_i\}$

    \ENDIF

\ENDFOR

\STATE $\hat{\mathcal{B}}\leftarrow\hat{\mathcal{B}}_0\cup\mathcal{A}$

\STATE For each candidate, $V_i=|\det J_f(\theta_i)|/q(\theta_i)$
\hfill\COMMENT{metric-volume weight}

\STATE $a\leftarrow
\dfrac{\displaystyle\sum_i V_i\,
\mathbf{1}\!\left[\min_{z_j\in\hat{\mathcal{B}}_0}\|z_i-z_j\|>r_{\mathrm{target}}\right]}
{\displaystyle\sum_i V_i}$
\hfill\COMMENT{estimated uncovered metric volume fraction}

\IF{$a<\varepsilon$}
    \STATE \textbf{break}
\ENDIF

\ENDFOR

\RETURN $\hat{\mathcal{B}}$

\end{algorithmic}
\end{algorithm}

The stopping criterion estimates the fraction of metric volume that remains uncovered by the current bank. For an exactly isometric mapping satisfying Eq.~\ref{eq:isometry}, the metric volume element satisfies

$$ \sqrt{|g(\theta)|}=|\det J_f(\theta)|. $$

Since the learned mapping is only approximately isometric, we use $|\det J_f(\theta)|$ as an approximate metric volume weight. For candidates drawn from a proposal distribution $q(\theta)$, the corresponding importance weight is

$$ w(\theta)=\frac{|\det J_f(\theta)|}{q(\theta)}. $$

The estimate is only as reliable as the effective sample size of the weights, i.e. how many candidates effectively contribute to the sum. Uniform proposals give an effective sample size of $\sim1\%$ of the batch in the 5D eccentric space, where the metric volume concentrates in a small region of low secondary mass, so a proposal approximating $\sqrt{|g|}$ (for example log-uniform in the masses) is needed for the criterion to be meaningful. The proposal affects only the order in which regions fill and the quality of this estimate: the acceptance rule is pure latent distance, so the resulting packing is valid for any $q$ with support covering $\Theta$.

After each batch, the uncovered volume fraction is estimated from that batch's own candidates: each candidate contributes $V_i$ to the denominator, and those farther than $r_{\mathrm{target}}$ from every template of $\hat{\mathcal{B}}_0$, the bank at the start of the batch, also contribute to the numerator. Placement terminates when this estimate falls below $\varepsilon$. The tolerance $\varepsilon$ does not correspond to a coverage guarantee: it controls how close placement runs to saturation, and the resulting coverage must be verified with an explicit injection campaign.

Placement at $r_{\rm target}$ constrains the distance between templates, whereas coverage is set by the distance from a signal to its nearest template, which is smaller: in the saturated bank the median signal-to-template distance is $0.64,r_{\rm target}$, so typical signals match well above the minimal match and the bank over-covers. Operating at the intended point instead requires templates spaced further apart, so that typical signals lie near $r_{\rm target}$. Reducing the bank to the size quoted in the main text raises the median signal-to-template distance to $1.07,r_{\rm target}$, the separation at which the match equals the minimal match.

\subsection{Further results}\label{sec:further_results}

\subsubsection{The template bank}

For the \textit{toy example}, the 3D aligned-spin space, we draw $10^4$ injections uniformly in the coordinates of the parameter space defined in Table~\ref{table:param_spaces}. The bank of $23{,}810$ templates covers $99.72 \pm 0.05\%$ of the injections, comparable to the $99.80 \pm 0.04\%$ achieved by a \texttt{mbank} stochastic bank of similar size ($23{,}167$ templates) on the same injections. The few uncovered injections (28 of $10^4$) cluster at large mass ratio and positive spin (median $q = 4.5$, $\chi = 0.66$). Coverage falls to $99.37 \pm 0.16\%$ for $\chi > 0.5$ and to $99.37 \pm 0.16\%$ for $M < 35,M_\odot$, against $99.8\%$ in the rest of the space. The worst-matched injection has $\mathrm{FF} = 0.911$, and only 6 injections fall below $\mathrm{FF} = 0.96$. As in the eccentric case, the template density is highest at low total mass and large mass ratio (Fig.~\ref{figure:bank_aligned}). This space is low-dimensional and well described by existing placement methods, so we use it as a control: \texttt{isobank} reproduces a bank of standard quality without any tuning specific to the space.

\begin{figure*}[h!]
	\centering
	\includegraphics[width=\textwidth]{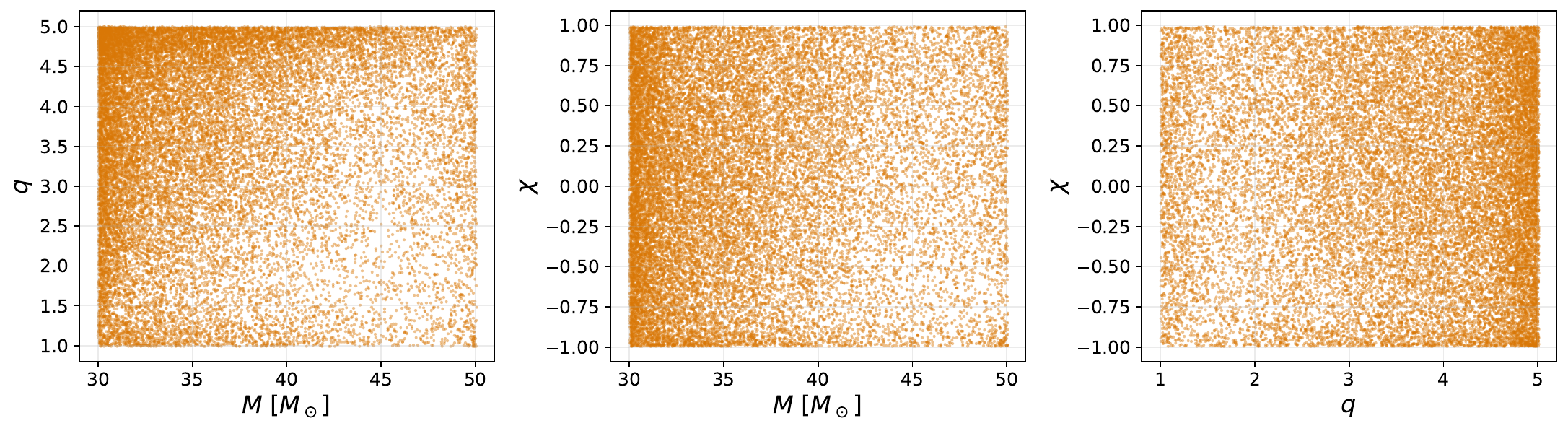} 
	\caption{Projections of the 3D aligned-spin template bank ($23{,}810$ templates, all shown) onto pairs of parameters. Template density is highest at low total mass and large mass ratio.}
	\label{figure:bank_aligned}
\end{figure*}

For the \textit{high-dimensional example}, the 5D eccentric space, we produce a simulated injection campaign (see definition in Section~\ref{sec:background}) of $10^4$ waveforms drawn uniformly in the coordinates of the parameter space defined in Table~\ref{table:param_spaces}. The resulting template bank gives an overall coverage of $96.6\%$ with $2.27$M templates. The uncovered injections are not uniformly distributed: coverage falls to $91.3 \pm 1.3\%$ for chirp masses $\mathcal{M} < 1.4,M_\odot$, to $94.5 \pm 0.4\%$ for mass ratios $q \geq 3$, and to $94.2 \pm 0.4\%$ for $\chi_1 < -0.3$, against $96.9\%$, $97.4\%$ and $97.9\%$ respectively over the rest of the space. Interestingly, the region $e \geq 0.12$ is covered at $96.5 \pm 0.4\%$, consistent with the $96.7 \pm 0.2\%$ found at lower eccentricity, so the high-eccentricity corner, which the training loss treats with a dedicated edge term, shows no loss of coverage. Fig.~\ref{figure:bank} shows the projected template density, which is strongly concentrated towards low component masses where the metric volume grows most rapidly.

\begin{figure*}[h!]
	\centering
	\includegraphics[width=\textwidth]{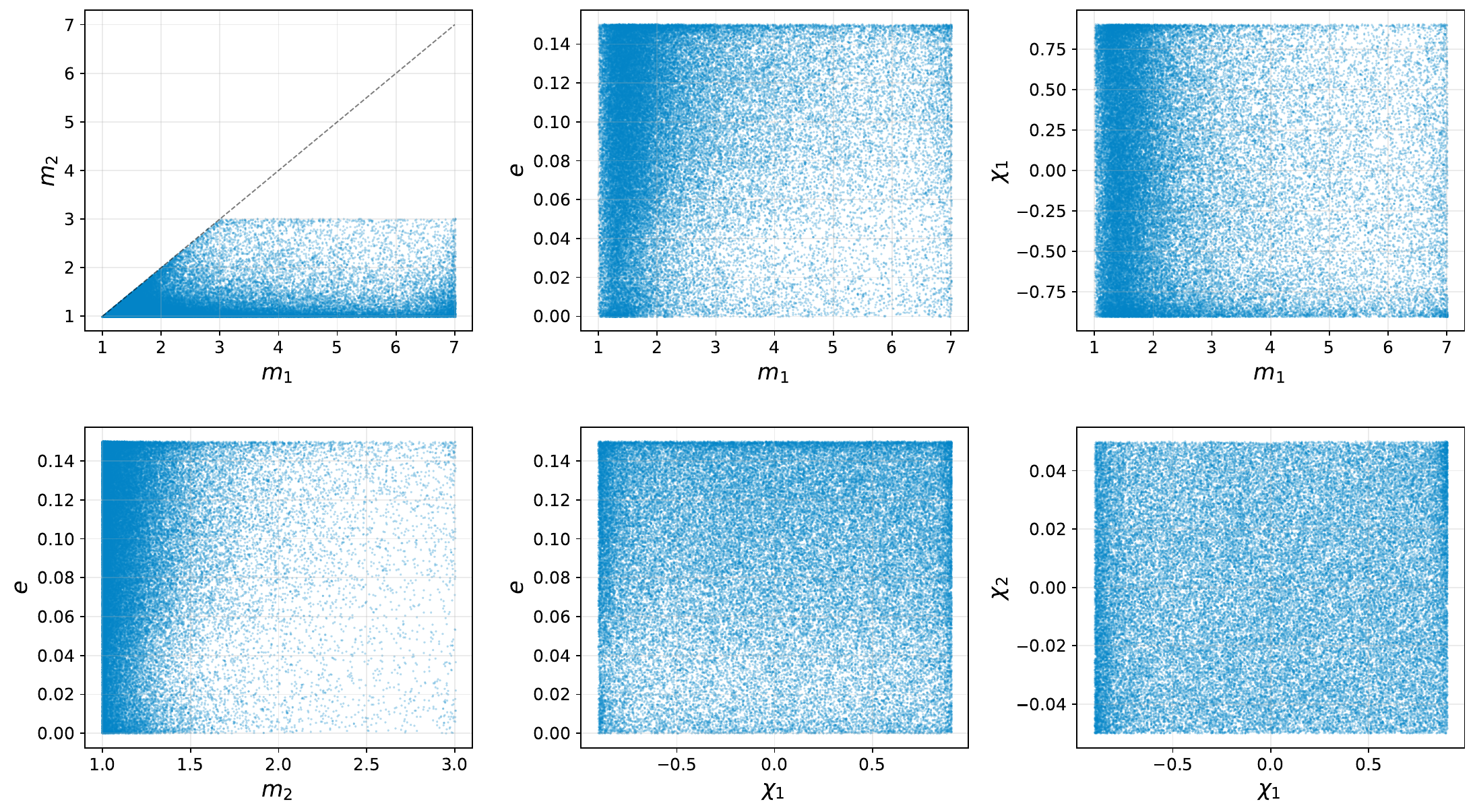} 
	\caption{Projections of the 5D eccentric template bank ($2{,}267{,}558$ templates) onto pairs of parameters. For readability, a random subset of $60{,}000$ templates is shown. Templates are concentrated at low component masses, where the metric volume density is largest, and are spread nearly uniformly in $\chi_1$, $\chi_2$ and $e$. The dashed line in the $m_1$--$m_2$ panel marks $m_1 = m_2$.}
	\label{figure:bank}
\end{figure*}
\subsubsection{The match}

As discussed in Section~\ref{sec:background}, the match $\mathcal{M}(\theta_1,\theta_2)$ measures the noise weighted overlap between the waveforms at two points in parameter space, maximized over time and phase, and is therefore the quantity that determines template bank coverage. The target mismatch $\mu = 0.03$ corresponds to a target match $\mathcal{M} = 0.97$.
To verify that the trained flow reproduces this geometry, we generate pairs of points $(\theta_1,\theta_2)$ separated by a range of proper distances and compare their Euclidean separation in the latent space, $d_z(\theta_1,\theta_2) = |z(\theta_1)-z(\theta_2)|$, with the true waveform match. Fig.~\ref{figure:match} shows that $d_z$ is a faithful surrogate for the waveform geometry on the scale relevant to template placement: pairs separated by $r_{\rm target}$ have a mean match consistent with the calibration target of $0.97$. Beyond this scale the relation broadens, but placement requires accurate distances only near $r_{\rm target}$. Larger separations enter only through the nearest-neighbour ranking used to shortlist candidate templates when evaluating fitting factors, where the ordering of distances matters rather than their exact values.

\begin{figure}[h]
    \centering
    \begin{minipage}[c]{0.60\columnwidth}
        \centering
        \includegraphics[width=\linewidth]{
            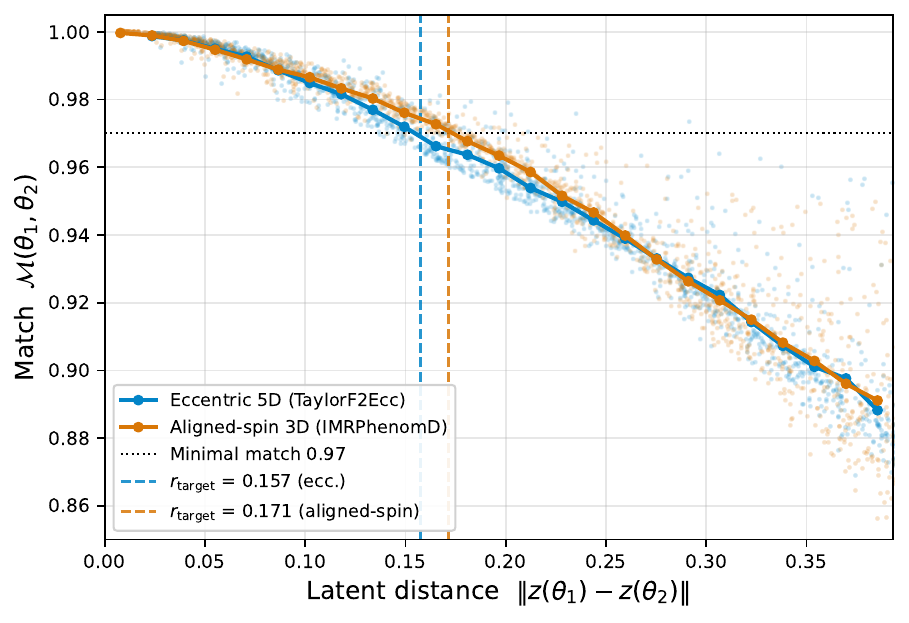
        }
    \end{minipage}
    \hfill
    \begin{minipage}[c]{0.39\columnwidth}
        \captionof{figure}{Individual pairs $(\theta_1,\theta_2)$ shown as points: the true waveform match $\mathcal{M}(\theta_1,\theta_2)$ against their Euclidean distance in the flow's latent space, $|z(\theta_1)-z(\theta_2)|$. Solid lines with markers are means in 25 equal-width bins. The horizontal dotted line marks the minimal match $1-\mu=0.97$; the vertical dashed lines mark each space's calibrated placement radius $r_{\rm target}$ (0.157 eccentric, 0.171 aligned-spin), the radius at which the mean mismatch equals $\mu=0.03$ and at which templates are packed in the latent space.
        }
        \label{figure:match}
    \end{minipage}
\end{figure}
\end{document}

%% file: math_commands.tex
\usepackage{amsmath,amsfonts,bm}

\def\eqref#1{equation~\ref{#1}}

\def\1{\bm{1}}

\DeclareMathAlphabet{\mathsfit}{\encodingdefault}{\sfdefault}{m}{sl}
\SetMathAlphabet{\mathsfit}{bold}{\encodingdefault}{\sfdefault}{bx}{n}

